\documentclass[aps,prb,groupedaddress,twocolumn]{revtex4}
\usepackage{graphicx}

\begin{document}
\title{Superconductor-insulator transition of Josephson-junction
arrays on a honeycomb lattice in a magnetic field}

\author{Enzo Granato}

\address{Laborat\'orio Associado de Sensores e Materiais,
Instituto Nacional de Pesquisas Espaciais, 12227-010 S\~ao Jos\'e dos
Campos, SP, Brazil}

\begin{abstract}
We study the superconductor to insulator transition at zero temperature
in a Josephson-junction array model on a honeycomb lattice with $f$
flux quantum per plaquette. The path integral representation of
the model corresponds to a (2+1)-dimensional classical model, which is
used to investigate the critical behavior by extensive Monte Carlo simulations
on large system sizes. In contrast to the model on a square lattice,
the transition is found to be first order for $f=1/3$  and continuous for $f=1/2$ but in  a different
universality class. The correlation-length critical exponent is estimated from
finite-size scaling of vortex correlations. The estimated universal conductivity at the transition
is approximately four times its value for $f=0$.  The results are compared with experimental observations
on ultrathin superconducting films with a triangular lattice of nanoholes
in a transverse magnetic field.

\end{abstract}

\maketitle

\section{Introduction}
Superconductor-insulator transitions in Josephson-junction
arrays are particularly interesting
as physical realizations of a quantum phase transition
\cite{fazio,sondhi,geerligs,vdzant,stroud,doniach,gk90,schon,eg04,cha1}.
These arrays also provide a simple model for studies of phase coherence
in inhomogeneous superconductors. Artificial arrays can be fabricated in different
geometries both in $1$ and $2$ dimensions (2D) \cite{vdzant} as a lattice of
superconducting grains coupled by the Josephson or proximity effect,
with well-controlled parameters. As a theoretical model, a Josephson-junction array
is closely related to the Bose-Hubbard model, where Cooper pairs interact on a
lattice potential, in the limit of large number of bosons
per site \cite{sondhi,cha1} and it also shows interesting analogies
with ultracold atoms on optical lattices \cite{nature,micnas}.

If charging effects due to the small capacitance of the grains are
sufficiently large, strong quantum fluctuations of the phase of
the superconducting order parameter drive the Josephson-junction array into an
insulating phase leading to a superconductor-insulator transition
for increasing ratio of charging energy to the Josephson coupling.
Under an applied magnetic field, frustration
effects lead to phase transitions with distinct universality classes, characterized
by different critical exponents and universal conductivities at the transition,
depending on the geometry of the array and the value of the frustration parameter $f$. The
parameter $f$ corresponds the number of flux quantum per plaquette.
For a Josephson-junction array model on a square lattice,
the superconductor-insulator transition is continuous at low-order rational frustrations,
$f=1/2, 1/3$ and $1/4$, and  the critical behavior have already
been investigated in detail numerically \cite{cha1,cha2,cha3,stroud08}, both in relation
to experiments \cite{geerligs,vdzant} and theoretical predictions
\cite{gk90,schon,fisher}. In particular, for $f=1/2$ the universal conductivity
at the transition \cite{gk90,cha2} is approximately two times its value for $f=0$.  For $f=1/5$,
the transition becomes first order \cite{cha1}.
Similar results are not available for Josephson-junction arrays on a honeycomb lattice,
despite the general interest on quantum phase transitions. In part, this is due to
the lack of experimental data for artificial arrays in this geometry.
Recently, however, there has been a growing interest in a related system, in  the form
of an ultra-thin superconducting film with a triangular lattice of
nanoholes \cite{valles07,valles11,valles13,kopnov,valles15}.
A simple model for this
system consists of a Josephson-junction array on a honeycomb lattice, with the triangular
lattice of nanoholes corresponding to the dual lattice, and it has already been used to
investigated the thermal resistive transition in absence of
charging effects \cite{eg13}. Nevertheless, quantum phase fluctuations should be taken into account
since the system undergoes a superconductor to insulator transition for decreasing film thickness.
The transition is very sensitive to low applied magnetic fields and is the
analog of the transition in a Josephson-junction array for increasing ratio of the
charging energy to the Josephson coupling at different frustration parameters.
In the insulating phase near the transition,
the magnetoresistance oscillates with the applied magnetic field at low
temperatures, displaying minima at integer values of $f=n$ and
maxima at $f=n+1/2$, corresponding to quantum phase transitions at
different critical points.

In the absence of magnetic field, $f=0$, or integer values of $f=n$,
the difference between periodic square and honeycomb lattices is irrelevant
to the critical behavior. However, for non integer values of $f$, commensurability
effects of the vortex lattice on the pinning centers may lead to transitions
which are in different universality classes. The $f=1/2$ case is of particular
interest since the topology of the honeycomb lattice leads to geometrical frustration,
which can destroy phase coherence even in the absence of charging effects.
For the array on a square or triangular lattice, the dual
lattice formed by the plaquette centers is bipartite and a
commensurate vortex lattice with density $1/2$ and double
degeneracy is possible \cite{teitel,shih85}.
For the honeycomb array, however, the dual lattice is triangular
and the resulting vortex lattice is incommensurate, with a
macroscopically degenerate ground state \cite{shih85}. Thermal fluctuations
leads to  unusual phase-coherence and vortex-order transitions
as a function of temperature, which has been investigated
both analytically \cite{korsh04,korsh12}
and numerically \cite{shih85,leeteitel,reid,eg12,eg13}
but are still not well understood.
However, the phase-coherence and vortex-order transitions at zero temperature for
increasing quantum phase fluctuations have not been investigated in detail.

In this work, we study the superconductor-insulator transition at zero temperature
using a self-charging model of Josephson-junction
arrays on a honeycomb lattice, with $f$ flux quantum per plaquette. The path integral representation of
the model corresponds to a (2+1)-dimensional classical model, which is
used to investigate the critical behavior by extensive Monte Carlo simulations on large
system sizes. A short account of some preliminary calculations for smaller system sizes and for the phase correlations
has been reported in a conference proceedings \cite{conf}.
In contrast to the results for the square lattice \cite{gk90,cha1,cha2,cha3},
we find that for the honeycomb lattice the transition
is first order for $f=1/3$  and continuous for $f=1/2$
but in a different universality class. Here, the first order transition is determined from
the finite-size behavior of the free-energy barrier between coexistent states and
the critical behavior for $f=1/2$ is determined from the finite-size scaling of vortex correlations.
We also obtain numerically the universal conductivity at the transition from a scaling analysis
of the phase stiffness \cite{cha2,cha3}.
The results indicate that the phase-coherence and vortex-order
transitions occur at the same critical point with a single divergent correlation length. In particular,
the universal conductivity at the transition is found to be about four times its value for $f=0$. The results are compared with experimental measurements on ultrathin superconducting films with a periodic triangular pattern of nanoholes\cite{valles07}. The magnetoresistance oscillations
and activation energy behavior observed in this system in the insulating phase at low temperatures and low
magnetic fields, are consistent with the Josephson-junction array model.
We argue that the absence of secondary minima at $f=1/3$ predicted by the model and the activation-energy behavior with film thickness are due to effects of Josephson-coupling disorder.

\section{Model and Monte Carlo simulation}
We consider a Josephson-junction array model on a honeycomb lattice,
as illustrated in Fig. 1, described by the Hamiltonian
\cite{fazio,doniach}
\begin{equation}
{\cal H} = -{{E_c}\over 2} \sum_i \left( {d \over { d \theta_i }}
\right) ^2 - \sum_{<ij>} E_{ij} \cos ( \theta_i -
\theta_{j}-A_{ij}). \label{hamilt}
\end{equation}
The first term in Eq. (\ref{hamilt}) describes quantum
fluctuations induced by the charging energy  $E_c = 4 e^2 /C$ of a
non-neutral superconducting grain located at site $i$, where $e$
is the electronic charge. The effective capacitance to the ground of each grain $C$
is assumed to be spatially uniform.
The second term in (\ref{hamilt}) is the
Josephson-junction coupling between nearest-neighbor grains
described by phase variables  $\theta_i$. In the present calculations we assume a spatially
uniform Josephson coupling,  $E_{ij}=E_J$.
The effect of the magnetic field $\bf B$ applied in the perpendicular ($\hat z$-direction)
appears through the link variables $A_{ij}=(2\pi/\Phi_o)\int_{r_i}^{r_{j}} \bf A \cdot d \bf r$,
where $\Phi_o=hc/2e$ is flux quantum and $\bf A$ is the vector potential. Since $\bf B = \nabla \times \bf A $,
the gauge-invariant sum around each elementary hexagonal
plaquette of the array is constrained by $\sum_{ij} A_{ij} = 2 \pi f$
with $f=\Phi/\Phi_o$.  The frustration parameter $f$ corresponds to the number of
flux quantum per hexagonal plaquette. The properties of the model are periodic in $f$ with period $f = 1$ and
have a reflection symmetry about $f=1/2$.

To study the quantum phase transition at zero temperature,
it is useful to employ the imaginary-time path-integral
formulation of the model \cite{sondhi}. In this representation, the
2D quantum model of Eq. (\ref{hamilt}) maps into a (2+1)D classical statistical
mechanics problem. The extra dimension corresponds to the
imaginary-time direction. Dividing the time axis $\tau$ into
slices $\Delta \tau$, the ground state energy
corresponds to the reduced free energy $F$ of the classical
model per time slice.
The classical reduced Hamiltonian can be written as
\begin{eqnarray}
H= &&-\frac{1}{g} [ \sum_{\tau,i}
\cos(\theta_{\tau,i}-\theta_{\tau+1,i}) \cr &&
+\sum_{<ij>,\tau} \cos(\theta_{\tau,i}-\theta_{\tau,j}-A_{ij}) ],
\label{chamilt}
\end{eqnarray}
where a re-scaling of the time slices has been performed in order to obtain
space-time isotropic couplings.
In this equation, $\tau$ label the sites in the imaginary-time direction.
The ratio $g =
(E_c/E_J)^{1/2}$, which drives the superconductor to insulator transition for the
model of Eq. (\ref{hamilt}), corresponds to an effective "temperature" in the
3D classical model of Eq. (\ref{chamilt}). The energy gap $\Delta$ of the insulating phase is related
to the phase correlation length in the time direction $\xi_\tau$ by $\Delta=1/\xi_\tau$.
The classical Hamiltonian of Eq. (\ref{chamilt}) can be viewed as an
XY model on a layered honeycomb lattice, where  frustration effects exist
only in the honeycomb layers.

\begin{figure}
\includegraphics[bb= -0cm 0cm  10cm 7cm, width=7.5cm]{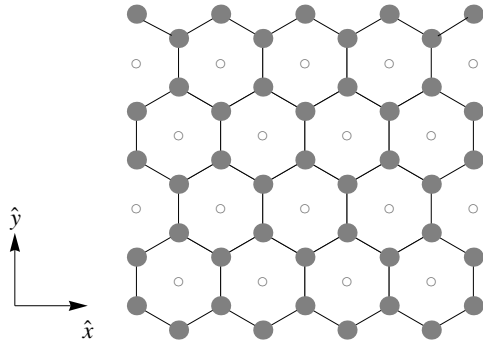}
\caption{(a) Josephson-junction array on a honeycomb lattice. Filled
circles represent superconducting grains and the lines the Josephson
junctions between them. Open circles represent the sites of the dual
triangular lattice. } \label{harray}
\end{figure}

\begin{figure}
\includegraphics[bb= -0cm 0cm  10cm  7cm, width=7.5cm]{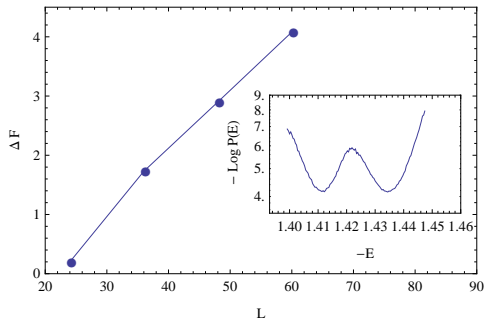}
\caption{Finite-size
dependence of the free-energy barrier $\Delta F $ for $f=1/3$. Inset: restricted free energy $A= -\ln p(E) $ as a function energy density $-E$ for system size $L=36$, near the transition.}
\label{histf13}
\end{figure}


MC simulations are carried out using the 3D classical Hamiltonian of Eq. (\ref{chamilt}).
The parallel tempering method \cite{nemoto} is used in the simulations
with periodic boundary conditions. The method is implemented using
a number of replicas of the system, each one with a different coupling $g$
within a range around the critical point. These replicas are
simulated in parallel and the corresponding configurations
are  exchanged with a probability distribution satisfying detailed balance.
For convenience, the honeycomb lattice is defined on a rectangular geometry
(Fig. \ref{harray}) with linear size given by a dimensionless length $L$.
In terms of $L$, the linear size in the $\hat x$  and $\hat y$ directions
correspond to $L_x=L\sqrt{3}$ and $L_y=\frac{3}{2}L$, respectively.
We choose a gauge where
$A_{ij}= 2 \pi f n_y/2$, on the (tilted) bonds along the rows in the
$\hat x$ direction numbered by the integer $n_y$ and $A_{ij}=0$ on the
bonds along the  $\hat y$ direction.
For the finite-size scaling analysis, calculations are done for different
values of $L$ with the linear size in the time direction $L_\tau$ set equal to linear size
in the spatial direction $L$. This choice assumes implicitly that the dynamic
critical exponent $z \sim 1$. More generally, a quantum phase transition
shows intrinsic anisotropic scaling, with  diverging correlation
lengths $\xi$ and $\xi_\tau$ in the spatial and time directions
\cite{sondhi}, respectively. They are related by the dynamic critical exponent $z$
as $\xi_\tau \propto \xi^z$. The scaling
behavior discussed in the next Section is indeed consistent with $z \sim 1$.
In particular, for  the unfrustrated cased, $f=0$, the transition is in  the universality class
of the 3D classical XY model, for which it is known that  $z=1$. For a Josephson-junction array on
square lattice \cite{cha1,cha2}, it has already been verified that
$z\sim 1$ is consistent with the scaling behavior in absence of disorder.
\section{Numerical results and discussion}

We first consider the histogram of the probability distribution $p(E)$ for the energy
density near the transition. The structure and finite-size dependence of $p(E)$
provide information on the nature of the transition, if it is first order or continuous \cite{LeeK}.
For $f=1/3$ (Inset of  Fig. \ref{histf13}), the restricted free energy defined as $A= -\ln p(E) $
shows a double minima structure. This suggests that there are two coexisting phases,
separated by a free-energy barrier $ \Delta F = A_M-A_m $,
where $A_m$ corresponds to one of the minima of $A(E)$
and $A_M$ corresponds to the maxima between them.  Fig. \ref{histf13} shows that $\Delta F (L)$
increases with system sizes indicating that the transition is first order.
On the other hand, the double minima feature is absent for
$f=1/2$ even for larger system sizes,
which is consistent with a continuous phase transition.
This is in sharp contrast with the results for the same model on a square lattice
\cite{cha1,cha2}, where the transition becomes first order at much smaller frustration $f=1/5$.
For $f=0$, the transition is also continuous and in the universality class of the
classical XY model in three dimensions, where the dynamic exponent is $z=1$ and
the correlation length exponent \cite{justin} is $\nu =0.671$.
However, for $f=1/2$ the critical properties are not known in detail. Some results were obtained previously for the phase-coherence transition \cite{conf}. Here,, we consider the scaling behavior of the vortex transition and determine the corresponding critical exponents. This allows us to consider the interesting question of the sequence of these transitions,
if they occur separately or at the same critical point as a function of the coupling $g$.



To investigate the  vortex-order transition, we study the scaling behavior of the  finite-size correlation
length given by  \cite{cooper}
\begin{equation}
\xi^v(L,g) = \frac{1}{2\sin(k_0/2)}[S(0)/S(k_0) - 1]^{1/2}.
\label{cdef}
\end{equation}
Here $S(k)$ is  the Fourier transform of the
vortex correlation function $C^v(r)$ and $k_0$ is the smallest nonzero
wave vector. This definition of finite-size correlation length
corresponds to  a finite-difference approximation to
the infinite system correlation length
$\xi^v(g)^2= -\frac{1}{S(k)} \frac{\partial S(k) }{\partial k^2} |_{k=0}$,
taking into account the lattice periodicity. If the phase transition is continuous,
$\xi^v(L,g)$ should satisfy the scaling form \cite{cooper}
\begin{equation}
\xi^v/L = F(L^{1/\nu} \delta g ),
\label{correl}
\end{equation}
where $F(x)$ is a scaling function. According to this scaling form, curves of  $\xi^v/L$ as a function of $g$,  should cross at the same critical coupling $g_c$, for different system sizes $L$. Moreover, a scaling plot of $\xi^v/L$ $\times$ $L^{1/\nu}\delta g$ sufficiently close to $g_c$ for different $L$ should collapse on to the same curve.

For $f=1/2$, each honeycomb layer of the 3D classical XY model in Eq. (\ref{chamilt}) is fully
frustrated with a highly degenerate ground state \cite{shih85}. This leads to low-energy states
for the layered model, which are  macroscopically degenerate. In this  case, the choice of an order parameter
to describe the correlation function $C^v(r)$ is not obvious since the ordered phase may correspond to
an aperiodic or glassy-like pattern.
Here we find convenient in the numerical simulations to define the correlation function in
terms of an overlap order parameter \cite{bhat}.
The overlap order parameter  of vortex variables is given by
$q_v(p)=v_p^1v_p^2$, where $v_p=n_p-f$ is the net vorticity at site $p$ of the dual lattice (Fig. \ref{harray}).  The vorticity  is defined as $n_p=\sum_{ij}(\theta_{\tau,i}-\theta_{\tau,j}-A_{ij})/2 \pi$,
where the summation is taken around the corresponding elementary hexagonal plaquette
and the gauge-invariant phase difference is restricted to the interval
$[-\pi, \pi]$. For the fully frustrated case, $f=1/2$, the vortex variables. or
chirality variables, $q_v=\pm 1/2$ are Ising-like variables  measuring the direction
of the circulating current in an elementary plaquette. The vortex correlation function in the
spatial direction is obtained as
\begin{equation}
C^v(r) =\frac{1}{N} \sum_{\tau,p} <v_{\tau,p} \  v_{\tau,p+r}>,
\label{cfunc}
\end{equation}
where $N$ is total number of dual sites. An analogous expression is
used for the correlation function $C^v_\tau(r)$ in the time direction.
Eq. (\ref{cdef}) is used to obtain the  finite-size correlation lengths
in the spatial and time directions from the corresponding correlations functions.
Note that the correlation length defined in terms of the overlap order
parameter  may have a different magnitude from the one defined in
terms of a single copy. However, they should display the same critical
behavior near the transition \cite{op}.



\begin{figure}
\includegraphics[bb= -0cm 0cm  10cm  7cm, width=7.5cm]{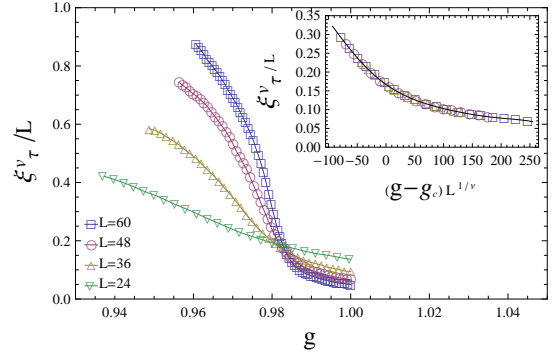}
\caption{  Vortex correlation length in the time direction
$\xi_{\tau}^v/L$ for $f=1/2$ and systems sizes $L$.
Inset: scaling plot for data near the transition and  $L \ge 36$
with $g_c= 0.9832$ and $\nu =0.40$. }
\label{corrvtau}
\end{figure}

\begin{figure}
\includegraphics[bb= -0cm 0cm  10cm  7cm, width=7.5cm]{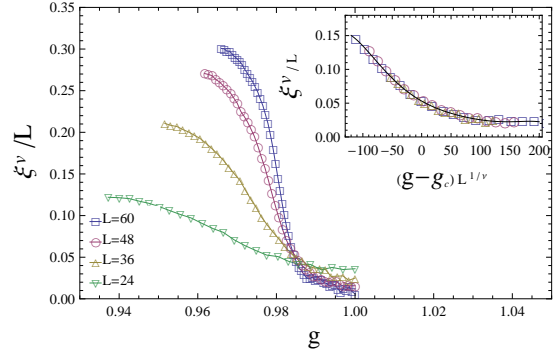}
\caption{ Vortex correlation length in the $\hat x$ spatial direction
$\xi^v/L$ for $f=1/2$.
Inset:  scaling plot for data near the transition and $L \ge 36$
 with $g_c=0.9844 $ and $\nu =0.40 $. }
\label{corrvy}
\end{figure}

In Figs. \ref{corrvtau} and \ref{corrvy} we show the finite-size behavior of the vortex correlation length
in the time and spatial direction, $\xi_{\tau}^v$ and $\xi^v$,  scaled by the system size $L$ as function of $g$.
The curves for the largest systems cross at the same point, providing evidence of  a continuous phase transition \cite{2d}.
In the insets of Figs. \ref{corrvtau} and \ref{corrvy}, we show a scaling plot of the data which verifies
the scaling form of Eq.  (\ref{correl}) and gives the estimates
$g_c= 0.9832(5)$ and $\nu= 0.40(5) $ from $\xi_{\tau}^v$, and  $g_c= 0.9844(5)$ and $\nu=0.40(5) $ from
$\xi^v$. The estimates of $g_c$ and $\nu$  are in reasonable agreement, within the errorbars, with the results obtained from
the phase correlation length \cite{conf},  $g_c= 0.9841(5)$ and $\nu= 0.48(4) $, suggesting that the superconductor-insulator transition is accompanied by a vortex disordering transition and described by a single divergent length scale.

\begin{figure}
\includegraphics[bb= -0cm 0cm  10cm  7cm, width=7.5cm]{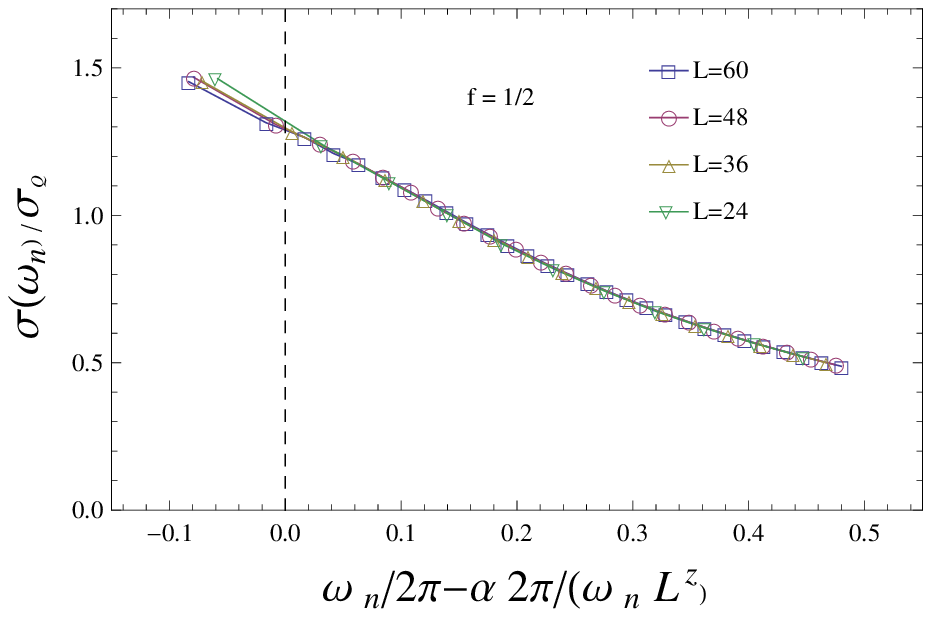}
\caption{Conductivity $\sigma(iw_n)$ at the critical coupling $g_c=0.9835$ for $f=1/2$ as
a function of the variable  $x=\frac{w_n}{2\pi} - \alpha \frac{2\pi}{w_n L_\tau}$, with
$\alpha=0.10$ and $z=1$. The universal conductivity  is given by the intercept with the $x=0$
dashed line for system sizes $L \ge 36 $, leading to  $\frac{\sigma^*}{\sigma_Q}= 1.29(2)$.}
\label{condf12}
\end{figure}

\begin{figure}
\includegraphics[bb= -0cm 0cm  10cm  7cm, width=7.5cm]{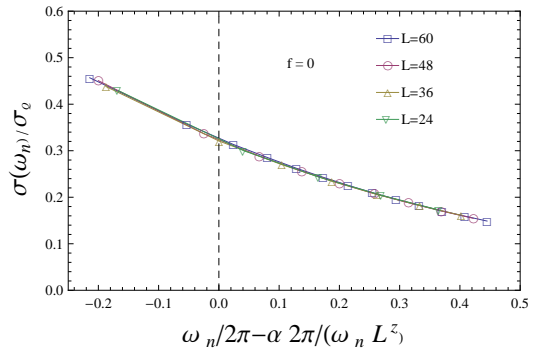}
\caption{Conductivity $\sigma(iw_n)$ at the critical coupling $g_c=1.725$ for $f=0$ as
a function of the variable  $x=\frac{w_n}{2\pi} - \alpha \frac{2\pi}{w_n L_\tau}$, with
$\alpha=0.25$ and $z=1$. The universal conductivity  is given by the intercept with the $x=0$
dashed line,   $\frac{\sigma^*}{\sigma_Q}= 0.32(2)$ }
\label{condf0}
\end{figure}

In addition to different critical exponents, the superconductor-insulator transition
for $f=1/2$ on a honeycomb lattice, is characterized by  a universal conductivity at the critical point,
which is significantly different from the square-lattice case. To determine its value we follow the
scaling method described by Cha {\it et al.}  \cite{cha2,cha3}.
The conductivity is given by the Kubo formula
\begin{equation}
\sigma = 2 \pi \sigma_Q \lim_{w_n\rightarrow 0}  \frac{\rho(i w_n)}{w_n},
\end{equation}
where $\sigma_Q=(2 e^2)/h$ is the quantum of conductance and $\rho(i w_n)$ is a frequency dependent phase stiffness
evaluated at the finite frequency $w_n=2 \pi n /L_\tau$, with $n$ an integer. The phase stiffness in
the $\hat x$ direction is given by
\begin{eqnarray}
\rho=&&\frac{1}{C N L_\tau  \  g}[ < \sum_{\tau,j} (\hat x \cdot \hat u_{j,j+\hat x} )^2 \cos(\theta_{\tau,j}
-\theta_{\tau,j+\hat x} \cr &&
- A_{j,j+\hat x})>
-\frac{1}{g} < |I(i w_n)|^2 > \cr &&
+\frac{1}{g} < |I(i w_n)| >^2],
\end{eqnarray}
where $C=4/3 \sqrt{3}$,  $N$ is the total number of sites in each layer,
\begin{eqnarray}
I(i w_n) =\sum_{\tau,j}(\hat x \cdot \hat u_{j,j+\hat x} ) \sin(\theta_{\tau,j} -\theta_{\tau,j+\hat x}- A_{j,j+\hat x}) e^{i w_n \tau},
\end{eqnarray}
and $\hat u_{j,j+\hat x}$ is a unit vector between nearest neighbors sites from $(\tau,j)$ to  $(\tau,j+\hat x)$.
At the transition, $\rho(i w_n)$
vanishes linearly with frequency and $\sigma$ assumes a universal value  $\sigma^*$, which can be  extracted from its
frequency and finite-size dependence \cite{cha2}
\begin{equation}
\frac{\sigma(iw_n)}{\sigma_Q} = \frac{\sigma*}{\sigma_Q}
- c (\frac{w_n}{2 \pi} - \alpha \frac{2 \pi}{w_n L_\tau}) \cdots
\label{cond}
\end{equation}
The parameter $\alpha$ is determined from  the best data collapse of the frequency
dependent curves for  different systems sizes  in a plot of $\frac{\sigma(iw_n)}{\sigma_Q}$ versus
$x=(\frac{w_n}{2\pi} - \alpha \frac{2\pi}{w_n L_\tau})$. The universal conductivity is
obtained from the intercept of these curves with the line $x=0$.
From the  scaling behavior of the conductivity for $f=1/2$ shown in
Fig. \ref{condf12} we obtain $\sigma^*/\sigma_Q= 1.29(2)$. In Fig. \ref{condf0} we show the
corresponding behavior for $f=0$, which gives $\sigma^*/\sigma_Q= 0.32(2)$. As would be expected, the
value of the universal conductivity for $f=0$ on a honeycomb lattice agrees with the known
result for the square lattice \cite{cha3}, $\sigma^*/\sigma_Q= 0.29(2)$, since the transitions are in
the same universality class.  However, the corresponding value for $f=1/2$ is significantly different.
Our estimate of the universal conductivity for $f=1/2$ on a honeycomb lattice is about four times its value for $f=0$ while
for the square lattice it is approximately \cite{cha2} two times. It is a clear evidence that, for $f=1/2$, these transitions
belong to different universality classes.

\begin{figure}
\includegraphics[bb= -0cm 0cm  10cm  6cm, width=7.5cm]{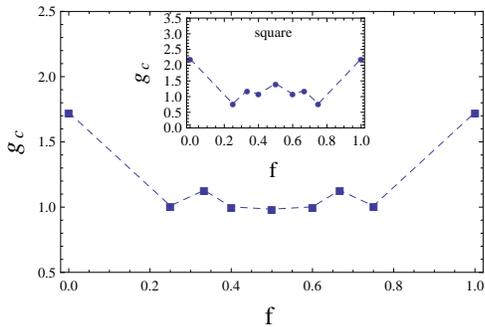}
\caption{ Critical coupling $g_c$ for the superconductor-insulator transition
on a honeycomb lattice, for different values of frustration $f$.
Inset: critical couplings for the square lattice.  }
\label{gcxf}
\end{figure}

\begin{figure}
\includegraphics[bb= -0cm 0cm  10cm  7cm, width=7.5cm]{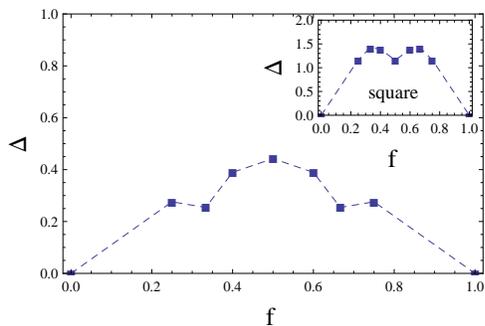}
\caption{ Energy gap $\Delta$ at $g_c=1.14$, near the superconductor-insulator transition
on a honeycomb lattice, for different values of frustration $f$.
Inset: energy gap $\Delta$ at $g_c=1.6$, near the transition on a square lattice. }
\label{delta}
\end{figure}

The critical couplings $g_c$ obtained numerically for different values of frustration are shown in
Fig. \ref{gcxf}. For comparison, we also show in the inset the critical
coupling obtained for the model defined on a square lattice. The lines connecting the data
are just guide to the eyes. $g_c(f)$ varies nonmonotonically with a large maximum at integer values of $f$ and a
secondary maximum at $f=1/3$. This is quite different from the behavior for the square
lattice, where a secondary maximum occurs at $f=1/2$.
The sharp variation of the critical coupling with frustration shows up in the behavior of the
energy gap $\Delta=1/\xi_\tau$ with frustration, near the transition. The phase correlation length
$\xi_\tau$ is obtained from Eq. (\ref{cdef}) using the corresponding correlation function in terms of
the phase variables $e^{i \theta_{\tau,j}}$, analogous to Eq. (\ref{cfunc}).
Since $g_c(1/2) < g_c(0)$, for a system with $g \lesssim g_c(0)$, the energy gap displays oscillations with deep minima at integer
values $f=n$ and maxima at $f=n+1/2$, as shown in Fig. \ref{delta}.  A secondary minima  at $f=n+1/3$ is also observed. In contrast, as shown in the Inset,  for the square lattice a secondary minima is expected at $f=n+1/2$.
For a continuous superconductor-insulator transition, the  gap of the insulating phase vanishes
as \cite{fisher} $\Delta \propto (g-g_c(f))^a$, with the exponent $a=z \nu$.
Given the above numerical estimates for the
critical exponents $\nu$ and $z$, the energy gap should vanish near the transition
with a power-law exponent $a=0.67$ at zero magnetic field and integer values of $f$, and
with $a=0.40$ at $f=n+1/2$. This sharp dependence on frustration determines the  magnetoresistance behavior of the Josephson-junction array in the insulating phase near the transition. In fact, at sufficient low
temperatures, the resistance is thermally activated $R\propto e ^{U/kT}$, and the activation energy $U$ corresponds to the gap of the insulating phase. Therefore the magnetoresistance should display oscillations with deep minima at integer
values of frustration, $f=n$, and maxima at $f=n+1/2$. Secondary minima at $f=n+1/3$
are also expected. In contrast, for the square lattice secondary minima are expected at $f=n+1/2$.

We now compare the expected behavior of the magnetoresistance with experimental observations
on ultrathin superconducting films with a triangular lattice of nanoholes\cite{valles07}
at low applied magnetic fields.
In the regime where phase fluctuations of the superconducting order parameter are more
important than amplitude fluctuations, \cite{cha1,giroud,emery} this system can be described by
an array of superconducting "grains" coupled by Josephson junctions in a suitable geometry.
The simplest model consists of a Josephson-junction array on a honeycomb lattice \cite{eg13}, with the
triangular lattice of nanoholes corresponding to the dual lattice (Fig. \ref{harray}),
which act as vortex pinning centers. The number of flux quantum per unit cell of the
triangular nanohole lattice corresponds to the frustration parameter $f$ of the honeycomb array. The superconductor
to insulator transition is tuned by decreasing the film thickness, which corresponds
to increasing  the coupling $g$ in the Josesphson-junction array model. In fact, measurements of
the resistance in the insulating phase near the transition show magnetoresistance oscillations
with the applied magnetic field, displaying minima at integer values of $f=n$ and
maxima at $f=n+1/2$, as expected. However, although the resistance is thermally activated,
the  activation energy increases roughly linearly with the deviation from the critical
value of the film thickness both for $f=0$ and $f=1/2$, which corresponds to an exponent $a \gtrsim 1$.
The secondary minimum at $f=1/3$ is not observed. This suggests that effects of quenched disorder
introduced by the fabrication process may be significant.
Disorder in the weak links between superconducting "grains" can arise from
inhomogeneities in the film thickness induced by the substrate \cite{valles11}. In fact,
it has been shown numerically \cite{eg13} that for the model of Eq. (\ref{hamilt}) in absence of charging effects, $E_c=0$,
disorder in Josephson-junction couplings $E_{ij}$ smooths out the secondary minima at $f=1/3$ near the thermally induced resistive transition. This kind of disorder should also affect the critical couplings $g_c(f)$ and and critical exponents $z$ and $\nu$ for the superconductor-insulator transition.

It turns out to be unfeasible to determine the effects of disorder  with the present Monte Carlo method
due to the long computer time required for averaging over many realization of disorder and large system sizes. As an alternative, we have performed additional simulations with a driven MC dynamics method \cite{eg04}. In this method, the layered honeycomb model of Eq. (\ref{chamilt}) is viewed as 3D superconductor  and the corresponding "current-voltage" scaling near the transition is used to determine the critical coupling and critical exponent \cite{wengel}. The numerical results \cite{egunp} show that indeed increasing disorder in $E_{ij}$ washes out the secondary minima at $f=1/3$ and lead to an  exponent $a \sim 1$, consistent with the experimental observation.

Geometrical disorder due to spatial irregularities of the system also leads to randomness in the  Josephson-coupling  but it has a more significant effect for increasing frustration. In fact, weak positional disorder of the grains or weak disorder in the plaquette areas \cite{gk86}, leads to disorder in  the magnetic flux per plaquette which increases with the field. Besides changing the universality class of the superconductor-insulator transition, it limits the number of oscillations in the magnetoresistance. This has already been observed in experiments on artificial Josephson-junction arrays with controlled amount of positional disorder\cite{benz88}, near the thermal resistive transition and in absence of charging effects. Very recently \cite{valles15}, it has also been
demonstrated in superconducting films with a pattern of nanoholes with controlled amount of positional disorder near the superconductor-insulator transition.

\section{Summary and Conclusions}

We have studied the superconductor to insulator transition in a self-charging model of
Josephson-junction arrays on a honeycomb lattice with $f$ flux quantum per plaquette.
From Monte Carlo simulations in the path-integral representation of the model,
we found that for $f=1/3$ the superconductor to insulator transition is
first order. For $f=1/2$, the transition is continuous and the
correlation length exponent and universal conductivity at the transition were estimated from finite-size
scaling. This is in contrast to the known results for the square lattice \cite{gk90,cha1,cha2,cha3},
for which the transition is also continuous for $f=1/3$. Moreover, the critical behavior
for $f=1/2$ is in a different universality class. In particular,
the universal conductivity is found to be about four times its value for $f=0$.
As for the square-lattice case \cite{vdzant}, it should be interesting to experimentally test
this prediction on artificially fabricated Josephson-junction arrays on a honeycomb lattice
with controlled parameters. It is interesting to note that the nature of the transitions for $
f=1/3$ and  $f=1/2$ on a honeycomb lattice is suggested from the ground state properties, in the absence of charging effects $E_c$.
For $f=1/3$, a vortex pattern with density $1/3$ commensurate with the triangular lattice (dual lattice),
with three-fold degeneracy is the ground state \cite{shih85}. If one neglects the coupling to phase variables, the vortex order should then be described by a three-state Potts model, which has a first-order transition in $3$ dimensions
\cite{janke}. Since the quantum phase transition can be described by a (2+1)-dimensional classical model, one thus should expect a first order transition for $f=1/3$. On the other hand, for $f=1/2$, similar arguments suggest that the vortex order should be described by an antiferromagnetic Ising model on a triangular lattice \cite{eg12}. This Ising model is geometrically frustrated and has a highly degenerate ground state but shows a continuous phase transition in $3$ dimensions, as a layered triangular lattice \cite{vojta}. In the present case for $f=1/2$, however, the transition is continuous but in a different universality class.
The behavior of the magnetoresistance oscillations observed experimentally in
superconducting films with a triangular lattice of nanoholes \cite{valles07,valles11,kopnov} are
qualitatively consistent with the predictions from the model. We argued that the absence of secondary minima
at $f=1/3$ predicted by the model and the approximately linear behavior of the activation energy with film thickness
are due to effects of Josephson-coupling disorder. The same system fabricated to be uniformly thick \cite{valles13} is not described by the present model, which assumes superconducting "grains" and weak links on a length scale of  nanohole size and  should belong to a different universality. The $f=1/2$ case is of particular interest since geometrical frustration
combined with thermal fluctuations leads to an unusual phase transition as a function of temperature \cite{eg13,eg12,korsh04,korsh12}. It should be of interest to investigate the effects of geometrical disorder
in the quantum transition of this system  \cite{gk86,stroud08,valles15}.

\acknowledgements

The author thanks J. M. Valles Jr.  for helpful discussions. This work was supported by  S\~ao Paulo Research Foundation (FAPESP, Grant \# 2014/15372-3 ) and computer facilities from CENAPAD-SP.

\medskip

Author contribution statement

The sole author had responsibility for all  parts of the manuscript.


\begin{thebibliography}{9}
\bibitem{2d} Similar calculations of the finite-size correlation length for different
system sizes as a function of temperature for the model of Eq. (\ref{hamilt}) without charging effects, $E_c=0$,
shows no crossing point or merging at finite temperature, suggesting that the critical
temperature of the thermal transition vanishes \cite{eg12}.
\end{thebibliography}
 \end{document}